\documentclass[fleqn,twoside]{article}
\usepackage[headings]{espcrc2}
\usepackage{amssymb}

\readRCS
$Id: espcrc2.tex,v 1.2 2004/02/24 11:22:11 spepping Exp $
\newcommand{\nn}{\nonumber}
\newcommand{\RR}{\mathbb R}
\newcommand{\vev}[1]{\left\langle#1\right\rangle}
\newcommand{\tsum}{\textstyle\sum}

\newcommand{\AmS}{{\protect\the\textfont2
  A\kern-.1667em\lower.5ex\hbox{M}\kern-.125emS}}

\title{A correspondence between $H_3^+$ WZW and Liouville theories on discs}
\author{Kazuo Hosomichi\address{\small\it
        Service de Physique Th\'eorique, CEA Saclay,
        F-91191 Gif-sur-Yvette Cedex, France}}
       
\runtitle{}
\runauthor{}

\begin{document}

\begin{abstract}
We discuss how the disc correlators of $H_3^+$ WZW model are
determined in terms of those of Liouville theory.
(A presentation based on a collaboration with S. Ribault \cite{Hosomichi-R}.)
\end{abstract}

\maketitle

Recently an interesting relation has been discovered between conformal
blocks of affine $SL(2)$ and Virasoro algebras.
By using it, the correlation functions of $H_3^+$ WZW model on sphere
have been related to those in Liouville theory \cite{Ribault-T}.

In this note we give a summary of the work \cite{Hosomichi-R}
which generalized this correspondence to disc (or the upper half plane)
and lead to an exact determination of disc correlators in $H_3^+$.

\vskip3mm

We begin by reviewing the correspondence for sphere correlators.
The space $H_3^+$ is a three dimensional hyperboloid in flat $\RR^{3,1}$,
\[
 x_0^2-x_1^2-x_2^2-x_3^2~=~ k\alpha' > 0,~~~  x_0>0.
\]
The space is coordinatized by a $2\times2$ hermite matrix $h$
with unit determinant and positive trace.
The CFT of interest is defined by the standard WZW action for
$h(z,\bar z)$ at level $k$, and has affine $SL(2)$ symmetry \cite{Gawedzki}.
We define the components $J^a(z)$ and $\bar J^a(\bar z)$ $(a=\pm,3)$
of the currents $-k\partial h h^{-1},~kh^{-1}\bar\partial h$
so that they are hermite conjugate to each other, and consider
correlators of ``$\mu$-basis'' primary fields $\Phi_j(\mu|z)$ which
have
\[
 L_0=\frac{-j(j+1)}{k-2},~~
 J_0^-=i\mu,~~
 \bar J_0^-=i\bar\mu.
\]
Their correlators satisfy the so-called KZ equations \cite{Knizhnik-Z}
that follow from the relation between the stress tensor and the currents.

We relate this model to Liouville theory with coupling
$b=(k-2)^{-\frac12}$ and central charge
\[
 c=1+6Q^2,~~~~
 Q\equiv b+1/b.
\]
The primary field $V_\alpha(z)$ with Liouville momentum $\alpha$
has conformal weight $\alpha(Q-\alpha)$.
The operator with $\alpha=-\frac{1}{2b}$ corresponds to a degenerate
representation of Virasoro algebra, and correlators containing
$V_{-1/2b}(z)$ are known to satisfy the so-called BPZ
equation\cite{Belavin-PZ} which is a differential equation of
second order in $z$.

An interesting relation \cite{Stoyanovsky} between KZ and BPZ equations
leads to a correspondence between $n$-point blocks of affine $SL(2)$ algebra
and $(2n-2)$-point Virasoro conformal blocks.
$n$ of their insertions are related by $\alpha_a=b(j_a+1)+\frac{1}{2b}$,
and the additional $(n-2)$ insertions on Virasoro side are all with
$\alpha=-\frac{1}{2b}$.
This was used in \cite{Ribault-T} to relate the correlators of
the two theories on sphere,
\begin{eqnarray}
\lefteqn{
 \vev{\prod_{a=1}^n\Phi_{j_a}(\mu_a|z_a)}
 ~=~ \delta^2(\tsum_{a=1}^n\mu_a)
 \cdot|\Theta|^\frac{1}{b^2}|u|^2
} \nn\\ &&
 \times
 \vev{\prod_{a=1}^nV_{\alpha_a}(z_a)\prod_{a=1}^{n-2}V_{-\frac{1}{2b}}(y_a)}.
\label{BPZKZ-s}
\end{eqnarray}
Here $u,~y_a$ and $\Theta$ are (implicitly) given by
\[
 \sum_{a=1}^n\frac{\mu_a}{\zeta-z_a}
 = u\frac{\prod_{a=1}^{n-2}(\zeta-y_a)}{\prod_{a=1}^n(\zeta-z_a)},
\]
\[
 \Theta =
 \prod_{a<\tilde a}(z_a-z_{\tilde a})
 \prod_{a<\tilde a}(y_a-y_{\tilde a})
 \prod_{a,\tilde a}(z_a-y_{\tilde a})^{-1}.
\]
Once verified for $n\le 3$, the correspondence (\ref{BPZKZ-s}) for
larger $n$ simply follows from the fact that the two sides behave
in the same manner under factorizations $n\to n_1+n_2$.

\vskip3mm

We wish to generalize this correspondence to correlators on discs.
Though there are several types of D-branes in $H_3^+$ WZW model known
(see e.g. \cite{Ponsot-ST}), here we focus on the one parameter family
defined by a plane
\begin{equation}
 x_3=\sqrt{k\alpha'}\sinh r,
\label{H2+}
\end{equation}
intersecting the hyperboloid.
Our convention is such that they are all characterized by
the boundary condition on currents,
$J^3=\bar J^3,~ J^\pm=\bar J^\pm$.
The system therefore has an unbroken affine $SL(2,\RR)$ symmetry,
and its zeromode part generates the isometry of the D-brane worldvolume
$H_2^+$.
We denote by $\Psi_l(\nu|w)$ the boundary operator between D-branes
(\ref{H2+}) with eigenvalues
\[
 L_0=-\frac{l(l+1)}{k-2},~J_0^-=i\nu.
\]
In Liouville theory, we consider boundary operators $B_\beta$ with
$L_0=\beta(Q-\beta)$ and FZZT boundary conditions \cite{Fateev-ZZ}
labelled by $s$.

\vskip3mm

To formulate the correspondence between disc correlators,
we first recall the relation between one-point
functions found in \cite{Ribault},
\begin{eqnarray*}
 \vev{\Phi_j(\mu|z)}_r \sim \vev{V_\alpha(z)}
 _{s=\frac{r}{2\pi b}-\frac{i}{2b}{\rm sgn}({\rm Im}\mu)},
\end{eqnarray*}
telling us how to relate the boundary conditions in two theories.
Now the correspondence between general disc correlators should take the form
\begin{eqnarray}
\lefteqn{
 \vev{\prod_{a=1}^n\Phi_{j_a}(\mu_a|z_a)
      \prod_{a=1}^m\Psi_{l_a}(\nu_a|w_a)}
} \nn\\
&=& \delta(\tsum_{a}(\mu_a+\bar\mu_a)+\tsum_a\nu_a)
    |u|\Theta^{\frac{1}{2b^2}}\cdot \nn\\
 &&\hskip-2mm \times
 \left<\prod_{a=1}^n V_{\alpha_a}(z_a)
       \prod_{a=1}^m B_{\beta _a}(w_a)
 \right. \nn\\&& \left.~~~~~~
       \prod_{a=1}^{n'} V_{-\frac{1}{2b}}(y_a)
       \prod_{a=1}^{m'} B_{-\frac{1}{2b}}(v_a)\right>,
\label{BPZKZ-d}
\end{eqnarray}
where $2n'+m'=2n+m-2$, and $(u,y_a, v_a)$ are determined via
a real function of $\zeta\in\RR$,
\begin{eqnarray*}
\varphi(\zeta) &\equiv&
 \sum_{a=1}^n\left(\frac{\mu_a}{\zeta-z_a}
                  +\frac{\bar\mu_a}{\zeta-\bar z_a}\right)
+\sum_{a=1}^m\frac{\nu_a}{\zeta-w_a}
 \nn\\
 &=& u\cdot\frac{\prod_{a=1}^{n'}|\zeta-y_a|^2\prod_{a=1}^{m'}(\zeta-v_a)}
                {\prod_{a=1}^n   |\zeta-z_a|^2\prod_{a=1}^m   (\zeta-w_a)}.
\end{eqnarray*}
$\Theta$ is a function of coordinate differences similar to the
case of sphere.
The most important part of the correspondence is how the boundary
conditions in two theories are related.
The boundary of the disc is cut at $\{w_a\}$ into segments
labelled by different $r$, and is cut further at $\{v_a\}$
in Liouville theory side.
Then the rule is that {\it a boundary segment labelled by $r$
in $H_3^+$ model maps to a Liouville FZZT brane with}
\[
 s~=~ \frac{r}{2\pi b}-\frac{i}{4b}{\rm sgn}\varphi(\zeta).
\]
Obviously $s$ is constant on each segment, as it should.
Our rule is also consistent with the fact in Liouville theory
that $s$ has to jump by $\pm\frac{i}{2b}$ where $B_{-1/2b}$
is inserted \cite{Fateev-ZZ}.

\vskip3mm

It is easy to show that our formula (\ref{BPZKZ-d}) behaves
consistently under different factorization limits, which actually
serves as a strong consistency check of our formula.
On the other hand, different factorization limits of a correlator of
$H_3^+$ model are usually not smoothly connected because of
transmutations of degenerate operators in Liouville theory along the way:
\[
 V_{-1/2b}(y_a) ~\longleftrightarrow~ B_{-1/2b}(v_a)B_{-1/2b}(v_{a'}).
\]
The construction of disc higher point functions from basic ones
is therefore not so straightforward in $H_3^+$ WZW model.
Our work \cite{Hosomichi-R} has argued that the disc correlators must
be continuous at transmutation points, and has shown that the formula
(\ref{BPZKZ-d}) is very likely the only solution to the continuity
as well as other consistency requirements.

The formula (\ref{BPZKZ-d}) has been used in \cite{Hosomichi-R} to
compute some disc structure constants of $H_3^+$ WZW model
which were not known before.
We have got the two point function of boundary operators
which is fully consistent with that of $N=2$ Liouville theory
\cite{Hosomichi}, and the bulk boundary propagator with the correct
semiclassical behavior at large $k$.
These serve as another non-trivial check of our formula.

\end{document}